\documentstyle[11pt,aaspp4,tighten]{article}

\def\arcs{\char'175\ }
\def\arcsc{\char'175 }

\def\etal{et~al.\ }

\def\hub{\ifmmode H_\circ\else H$_\circ$\fi}
\def\kms{~km~s$^{-1}$\ }
\def\dug{$\rm ^o~$}
\begin{document}

\singlespace

\title{STARBURSTS VERSUS TRUNCATED STAR FORMATION IN NEARBY CLUSTERS OF GALAXIES}

\author{James A. Rose\altaffilmark{1}}
\affil{Department of Physics and Astronomy, University of North Carolina, Chapel
Hill, NC 27599}
\affil{Electronic mail: jim@physics.unc.edu}

\author{Alejandro E. Gaba}
\affil{Department of Physics and Astronomy, University of North Carolina, Chapel
Hill, NC 27599}
\affil{Electronic mail: gaba@physics.unc.edu}

\author{Nelson Caldwell\altaffilmark{1}}
\affil{F.L. Whipple Observatory, Smithsonian Institution, Box 97, Amado AZ
85645}
\affil{Electronic mail: caldwell@flwo99.sao.arizona.edu}

\author{Brian Chaboyer\altaffilmark{1}}
\affil{Department of Physics and Astronomy, Dartmouth College, Hanover, NH 03755-3528}
\affil{Electronic mail: Brian.Chaboyer@Dartmouth.edu}
\altaffiltext{1}{Visiting Astronomer, Cerro Tololo Inter-American Observatory,
National Optical Astronomy Observatories, operated by the Association of
Universities for Research in Astronomy, Inc., under contract with the National
Science Foundation}

\begin{abstract}

We present long-slit spectroscopy, B and R bandpass imaging, and 21 cm 
observations of a sample of early-type galaxies in nearby clusters which
are known to be either in a star-forming phase
or to have had star formation which recently terminated.  From the long-slit 
spectra, obtained
with the Blanco 4-m telescope, we find that emission lines in the
star-forming cluster galaxies are significantly more centrally concentrated
than in a sample of field galaxies.  The broadband imaging
reveals that two currently
star-forming early-type galaxies in the Pegasus I cluster
have blue nuclei, again indicating that
recent star formation has been concentrated.  In contrast, the two galaxies for 
which star formation has already ended show no central color gradient.
The Pegasus I galaxy with the most evident signs of ongoing star formation
(NGC7648), 
exhibits signatures of a tidal encounter.  Neutral hydrogen observations 
of that galaxy with
the Arecibo radiotelescope reveal the presence of $\sim$4 x $10^8 M_{\sun}$
of HI.  Arecibo observations of other current or recent star-forming 
early-type galaxies in Pegasus I indicate smaller amounts of gas
in one of them, and only upper limits in others.  

The observations presented above indicate that NGC7648 in the Pegasus I
cluster owes its present star formation episode to some form of tidal 
interaction.  The same may be true for the other galaxies with centralized
star formation, but we cannot rule out the possibility that their outer disks
have been removed via ram pressure stripping, followed by rapid quenching of
star formation in the central region.

\end{abstract}


\section{Introduction}

It is now well-established that the star formation rate (SFR) in rich clusters of 
galaxies has declined markedly since z$\sim$0.5.  Evidence for this decline
was first established by Butcher \& Oemler (1978, 1984; see Margoniner \& de
Carvalho 2000 and references therein for recent results), who found a systematic
increase in the fraction of blue galaxies in rich clusters with
increasing redshift, a phenomenon known as the Butcher-Oemler effect.  {\it HST}
images have since shown
that the increase in the population of blue galaxies with z is also accompanied
by an increase in the fraction of spiral and interacting/merging galaxies.  Thus
the decline in SFR in rich clusters since z=0.5 appears connected to the
decrease in the population of spirals, and consequent increase in S0's
(e.g., Dressler \etal 1997; Poggianti \etal 1999), though the
details are complex (Fabricant \etal 2000).  The aim of this paper is to better understand what are
the driving mechanisms that produce the observed falloff in SFR.

While the basic observation of the Butcher-Oemler
effect appears to be generally well-accepted, there is much current debate
concerning the causes of the declining SFR, and to what
extent the rich cluster environment is involved.  On the one hand,
the cluster environment is certainly capable of both provoking and
quenching star formation in the member galaxies, since both gas removal 
mechanisms (e.g., Gunn \& Gott 1972; Abadi, Moore, 
\& Bower 1999) and tidal perturbations (e.g., Lavery \& Henry 1988;
Byrd \& Valtonen 1990; Moore \etal 1996, 1998; Bekki 1999), which can lead
to inward flow of gas and subsequent enhanced star formation, have been
proposed.  

Evidence for
enhanced star formation comes from the discovery of galaxies in 
distant clusters with strong Balmer absorption lines
and no emission lines in their spectra (e.g., Dressler \& Gunn 1983).  
This so-called ``E + A'' or ``K + A''
phenomenon is often attributed to the effect of a recent starburst
(e.g., Barger et al. 1996), thus indicating that cluster galaxies are
ending their star-forming careers in a final burst.  Recently, early-type
galaxies in {\it nearby} clusters, such as Coma, with similar K+A spectra
have been discovered with surprising frequency by Caldwell \etal (1993)
and Caldwell \& Rose (1997).  Subsequent long-slit spectroscopy (Caldwell
\etal 1996) and {\it HST} imaging (Caldwell, Rose, \& Dendy 1999) have indicated that
the recent star formation in these galaxies has been centrally concentrated
to the inner $\sim$2 kpc in radius.  In addition, models of the integrated 
spectra of some of these systems indicate that the recently terminated star
formation was burst-like, not simply a truncation of ongoing star 
formation (Caldwell \etal 1996; Caldwell \& Rose 1998). Thus in these cases 
the evidence favors a centrally concentrated star formation episode, 
rather than truncated disk star formation.

On the other hand, the identification of the K+A phenomenon with a
starburst episode has been challenged by Newberry, Boroson, \& Kirshner (1990) and by
Balogh \etal (1999).  These authors have pointed out that the distribution of
emission line and Balmer line strengths in cluster galaxies does not 
demonstrably differ from what one might expect for a collection of fading
spirals whose star formation was truncated rather suddenly.  Furthermore,
Balogh \etal (1999) propose that the fraction of K+A galaxies in distant
clusters is actually far lower (only about 1\% of their sample) than
previously indicated, and that there is no indication that the fraction of
K+A's exceeds that in the field, nor that the fraction has evolved significantly
from z=0.5 to the present epoch.  This result, based on the CNOC sample, 
contrasts strongly with the results of the MORPHS sample, which is reported
in Dressler \etal (1999) and Poggianti \etal (1999).  One problem is
that, except for immediately after a burst episode when the colors are very blue
and the
Balmer lines are very strongly enhanced, 
a fading starburst and a fading, truncated disk produce very similar spectra.
Thus exceptional care (high signal-to-noise) and fortune 
are required to distinguish the two.
As a result, there is 
considerable uncertainty as to whether the rather widely assumed poststarburst
picture for K+A galaxies is indeed correct, rather than the 
less dramatic situation of a truncated spiral.

A key unresolved question is the matter of the rapid evolution in
SFR since z=0.5: Is it  caused by final large bursts of star 
formation in cluster galaxies, or is it instead produced by a less dramatic
truncation of star formation?   Previous efforts
to distinguish between the two scenarios in the first question have largely
concentrated on interpreting the star formation histories of galaxies through
their colors and spectra.  However, another
way to resolve the issue
is by investigating the {\it spatial distribution of the last star formation}
in these galaxies.  If the primary effect is simply termination of star 
formation in an otherwise normal spiral disk, then we would expect to see the
fading of a roughly uniformly blue galaxy, with the exception, perhaps, of a redder
bulge.  If the mechanism is somehow related to starbursts, however, one would
expect the fading blue population to be centrally concentrated, since current
hypotheses for the production of starbursts all utilize variations
on a theme of tidal gravitational perturbations, which cause a pileup of gas in
the centers of galaxies, and hence a central starburst.

In this paper we have combined spectroscopic and imaging
data of star forming galaxies in nearby clusters to
evaluate whether recent star formation in
cluster galaxies has been centrally concentrated.
We place particular emphasis on the perspective of nearby clusters, since while
in distant clusters the processes producing the rapid evolution in SFR are more
in evidence, in nearby clusters one can study the processes with 
higher spatial resolution and higher S/N ratio, both in spectroscopy and
imaging.  Two datasets relevant to unusual star formation in nearby
clusters are presented in this paper.  First, we investigate the
spatial distribution of ionized gas in 10 early-type galaxies in nearby
rich clusters.  We compare the central concentration of the ionized gas with
that in an extensive sample of low-redshift field galaxies.  
The purpose is to evaluate
whether or not star formation in the cluster galaxies is unusually centrally
concentrated when compared with the field sample.  Second, we present B and R
images and HI observations
of four early-type galaxies in the Pegasus I cluster for which Vigroux \etal 
(1989) have previously found evidence for ongoing or recently completed star
formation.  The Pegasus I cluster is particularly relevant here
because of its low cluster x-ray emission, which indicates that ram pressure
stripping is not a significant process in this cluster.  Thus we have an 
unusual opportunity to isolate the effects of external (tidal) 
perturbations on the star formation histories in cluster galaxies.

The basic plan for this paper is as follows.
In \S 2 the imaging and spectroscopic data 
are summarized.  In \S 3 we compare the spatial
distribution of ionized gas in early-type galaxies in nearby rich clusters with 
that of the sample of ``field'' galaxies.  In \S 4 we discuss
radial luminosity profiles, morphology, and $B-R$ radial
color maps of four early-type galaxies in the nearby Pegasus I cluster, as well
as 21 cm observations of these galaxies. In \S 5 we
combine the results of the previous sections to assess the nature of the last
star formation in cluster galaxies, and place our results in the context of
other recent analyses.

\section{Observational Data}

\subsection{Spectroscopy}

Long-slit spectra were obtained of 10 galaxies in three nearby rich clusters. All
10 galaxies were classified to have early-type morphologies by Dressler (1980),
but were also known to have emission lines based on previous multi-fiber 
spectroscopy reported in Caldwell \etal (1993) and Caldwell \& Rose (1997).  The
long-slit spectra of five galaxies in the cluster DC2048-52 and of three 
galaxies in the double cluster DC0326-53/0329-52 (=A3125/A3128) were obtained
with the CTIO 4-meter telescope with the RC spectrograph and Loral 3K x 1K CCD
in August 1998.  The spectra cover the wavelength region $\lambda\lambda5000 -
8000$ \AA, at a dispersion of 1 \AA/pixel.  Thus the H$\alpha$, 
[NII]$\lambda\lambda6548,6584$, and [SII]$\lambda\lambda6717,6731$ emission 
lines are well centered in these spectra.  The scale along the slit direction is
0.5\arcsc/pixel, and the slit width is 1.7\arcsc. Exposure times were 20 minutes,
with repeat exposures taken in several cases.
Two other spectra, of the early-type emission-line galaxies
D15 and D45 in the Coma cluster, were obtained with the MMT, and have been
previously reported in Caldwell \etal (1996).  Briefly, these spectra also
cover the same emission lines as the CTIO 4-m spectra, at a dispersion of
0.8 \AA/pixel, and a scale along the slit of 0.3\arcsc/pixel.

Since the above long-slit spectra are used in this paper to explore the spatial 
distribution of star formation in emission-line galaxies in nearby clusters, a
reference sample of long-slit spectra of ``normal'' galaxies is required to use
as a benchmark for the cluster galaxies.  Recently, Jansen \etal (2000a,b) have
observed a large survey of relatively isolated galaxies, called the Nearby Field
Galaxy Survey (NFGS),
covering a wide range in absolute magnitude and morphology.  In addition to
broadband imaging (for luminosity profiles and colors) and globally averaged
spectroscopy (for use in comparison with spectroscopy of distant galaxies), they
have also obtained long-slit spectra, with the Tillinghast 1.5-m telescope of
the F. L. Whipple Observatory, for their entire sample of 198 galaxies.
The spectra cover the same [NII], [SII], and H$\alpha$ emission lines as our
spectra of the cluster galaxies, at a dispersion of $\sim$1.5 \AA/pixel, a
slit width of 3\arcsc, and a scale along the slit of 2.0\arcsc/pixel.  Exposure
times were typically 15-20 minutes.  While
the scale along the slit is at four times lower resolution than for the cluster
galaxies, the galaxies in the NFGS are typically 2-6 times closer than the
cluster galaxies.  As a result, after appropriate corrections are made to
spatially smooth the NFGS spectra to the same effective seeing as the cluster
galaxies, the two samples can be intercompared.  It was assumed
that the seeing was 1\arcs for both cluster and NFGS observations.  Moreover,
the majority of the cluster sample is at a redshift of 14000 \kms.  Thus to
bring a galaxy in the NFGS sample to the same effective spatial resolution as
the cluster galaxies, a smoothing was applied with a $\sigma$ of:

  \[  \sigma^2 = (\frac{14000}{4.708 V_r})^2 - (\frac{1.0}{4.708})^2, \]

\noindent where $V_r$ is the radial velocity of the NFGS galaxy (corrected to 
the Local Group velocity centroid, as given in Jansen \etal 2000a), and the 
factor $4.708$ accounts for the
2\arcsc/pixel binning of the NFGS spectra and the conversion from FWHM to 
$\sigma$.  

\subsection{Imaging of Pegasus I Cluster}

Images of four early-type galaxies in the Pegasus I cluster were obtained in
the B and R passbands using the 0.9-m telescope at CTIO on the nights of
September 3 -- 7, 1991.  The detector used was a Tektronix 1024$^2$ pixel CCD.
The scale at the focal plane of the 0.9-m telescope is 0.4\arcs per 24 $\micron$m 
pixel.  All observations were taken through thick and variable cloud cover, typically
1 -- 2 magnitudes.  Thus all color information is on a relative scale.  
Exposure times were typically 600 seconds each.  Multiple exposures were 
recorded in both B and R for all four galaxies, but due to the variable cloud
cover, the S/N ratio in the images varied greatly from one exposure to the
next.  Thus for most galaxies, a single exposure in each filter typically
contained the most information.  The seeing ranged from 1.4\arcs to
2.0\arcs FWHM, with the seeing in the R passband slightly better than in B.

Analysis of the images was accomplished using the IRAF package.  After the
frames were trimmed, bias-subtracted, and flat-fielded, a majority of the
cosmic rays in each image was removed using the 'cosmicrays' routine.  The
remaining cosmic rays in the vicinity of the galaxy were removed manually 
using the 'imedit' routine.  An average value of the sky was evaluated 
from statistics in sky-free zones well away from the galaxy.  The task
'ellipse' in IRAF was then used to obtain isophote fits for each galaxy
from the sky-subtracted images.  Radial luminosity profiles could then be
plotted from the isophote fits.  In addition, a model of the isophote fit
was contructed using the routine 'bmodel'.  This fit was 
subtracted from the original galaxy image, to 
look for fine structure in the image where the luminosity gradient is steep.

To produce color maps of the galaxies all selected B and R images of each object
were registered by selecting reference stars in the individual frames and using
the 'imalign' routine to determine centering offsets.  The image quality on each
frame was then determined using 'fitpsf' on selected stars in each of the 
registered frames and averaging the calculated FWHM's.  In general, the better
seing R frames had to be degraded to match the seeing on the B frames.  This
was done by gaussian-smoothing the R frames.  To obtain the radial color
profile, the final B and R images were fit using 'ellipse' (see above), and
the color profile was then determined from a point-to-point subtraction
of these profiles in magnitude form.  
Two types of errors have been considered for the luminosity and color profiles.
Statistical errors for the magnitudes were calculated by the 'ellipse' task
from the rms scatter of intensity data along the fitted ellipse.  We also
separately calculated the error in the magnitudes based on estimated
uncertainties in the sky subtraction, which dominates the errors at large
radial distances form the galaxy centers. Specifically, we assumed a 1\%
error in the sky background.  Both error estimates have been
plotted on the luminosity and color profiles. 

\subsection{HI Observations of Pegasus I}

Neutral hydrogen observations in the 21 cm transition were obtained for the 
four early-type galaxies
in the Pegasus I cluster that were also imaged in B and R.  The observations 
were acquired on July 14-15, 1999 using the 305-m Arecibo radiotelescope of the
National Astronomy and Ionosphere Center\footnote{The National Astronomy and
Ionosphere Center is operated by Cornell University under a cooperative
agreement with the National Aeronautics and Space Administration}.  The
L-narrow receiver was used at the upgraded Gregorian feed.  All four
subcorrelators covered a frequency range of 25 MHz with 1024 spectral
channels (at 5 \kms resolution), centered at 1403 MHz (i.e.,
centered at a redshift of 3675 \kms), and observations were recorded in both
circular polarizations.  The system temperature was approximately 30 K, and the
gain $\sim$10 K/Jy.  The observations consisted of sets of five minute
on-source and five minute off-source integrations (``scans'').  Each
five minute observation actually consisted of a series of five one minute
integrations (``dumps''), with each one minute observation separately recorded.
For NGC7557, NGC7611, NGC7617, and NGC7648, the total number of five-minute
scans is 5, 4, 6, and 5 respectively.  Hence, for instance, 25 minutes
of integration were carried out for NGC7648 on-source, and 25 minutes 
off-source.

The 21 cm data were reduced using the Analyz software at Arecibo. As a
first step, all 5 on and off dumps of 
each scan were added together. 
Then an ON/OFF-1 spectrum was obtained, to
subtract the background and normalize the spectrum to the correlator 
response, thereby yielding
percent of system temperature vs. frequency. The frequency was then converted to
heliocentric velocity. The system temperature was corrected for gain versus
zenith angle and temperature versus zenith angle dependencies, using available
calibration curves,
and converted to flux in Janskys. An average spectrum for each galaxy 
was obtained by averaging the ON/OFF-1 spectra of all 5 minute scans. 
Finally, the baseline
for each averaged spectrum was fit by a polynomial and 
subtracted. For the two detected galaxies the integrated 21 cm flux in Jy \kms 
was obtained. The total flux was converted to HI content in solar masses
using the relation found in, e.g., Binney \& Merrifield (1998)

\[ \frac{M_{HI}}{M_{\sun}} = 2.356 \times 10^5 \left( \frac{D}{Mpc}\right) ^2 \frac{\int_{-\infty}^{\infty}S(v)\,dv}{Jy~km~s^{-1}}, \]

\noindent where D is the distance to the galaxy and $S(v)$ is the flux in Jy as 
a function of velocity.

Using the absolute magnitudes in the B band given in Vigroux,
Boulade, \& Rose (1989), we calculated the total B band luminosity for each
galaxy. We then formed the HI mass to B band luminosity ratio, in solar units.

For the non-detected galaxies upper limits limits were calculated in the
following manner. 
The flux in 400 \kms wide bandpasses 
(the expected velocity widths for the galaxies)
were integrated from one side of the baseline-subtracted spectrum to the other,
centered at 200 km/s intervals.  The mean and standard deviation of the
distribution were calculated and the upper limit taken at the 3$\sigma$ level.
We also took the rms of the baseline-subtracted spectrum, 
multiplied by the square root of the number of channels in the above velocity
passband, and used 3 times this value as another estimate of the 3$\sigma$
upper limit.  In most cases the two estimates agreed closely, but in a few
cases, where the
noise in the spectra contained structure (due to radio frequency interference
and standing waves from the Sun in observations taken near sunrise), we 
considered the first method to provide a more conservative upper limit.

\section{Emission-Line Early-Type Galaxies in Nearby Clusters}

Having discussed the new observations, we now employ that data to establish
the centrally concentrated nature of star formation in our cluster galaxies.

As was mentioned in \S 1, multi-fiber spectroscopy of early-type galaxies in
five nearby rich clusters has revealed that $\sim$15\% of these galaxies have
experienced an unusual star formation history in the recent past (Caldwell
\etal 1993; Caldwell \& Rose 1997).  Since the galaxies studied have been
classified as E or S0 or S0/a by Dressler (1980) on high-quality photographic
plates, {\it any} recent star formation is in principle unusual.  In fact, 
(particularly in the SW region of the Coma cluster) many of these galaxies
have unusual star formation histories, regardless of morphology. That is, they
exhibit the K+A pattern of strong Balmer absorption lines but no emission, along
with relatively red colors, that is characteristic of galaxies which have
recently completed star formation, whether in a burst or through sudden
truncation of ongoing star formation.  Since K+A galaxies are rare in the
field at the present epoch, for {\it any} morphological type, we can indeed
conclude that the cluster K+A's are unusual.

The situation of cluster early-type galaxies with {\it current} star formation, as
evidenced by emission spectra characteristic of HII regions, is less certain.
Elliptical and S0 galaxies in the field with ongoing star formation are rare.
For example, 44 galaxies in the NFGS within the absolute magnitude range
-17.4$>$M$_B$$>$-22.0 are classified as early-type (i.e., with morphological
type parameter T$\le$0).  Of these, 22 are not detected in emission,
19 have nuclear emission that is dominated by an active galactic 
nucleus\footnote{We use the emission-line ratio [NII]$\lambda$6584/H$\alpha$$>$0.5 as a dividing line between AGN/LINER spectra and HII region/star formation
spectra, and when available, [OIII]$\lambda$5007/H$\beta$$>$0.5.  This dividing 
line is based on the distribution of AGN, LINERs, and HII regions in Fig. 5 of
Baldwin, Phillips, \& Terlevich (1981).} (AGN), 
and only 3 have nuclear emission dominated by star formation.
The latter 3 galaxies, A11352+3536, A12001+6439, and A22551+1931N, have been
categorized as a Markarian blue compact dwarf, a blue compact dwarf, and
a ``multiple galaxy'' respectively (as reported in NED), and hence are unusual
for their early-type classifications.
Thus in principle, the early-type emission-line galaxies found in clusters are 
truly experiencing unusual star formation.  {\it However, one must also consider the
possibility that these galaxies are simply misclassified spirals, in which
case ongoing star formation is to be expected.}  

\begin{figure}
\plotone{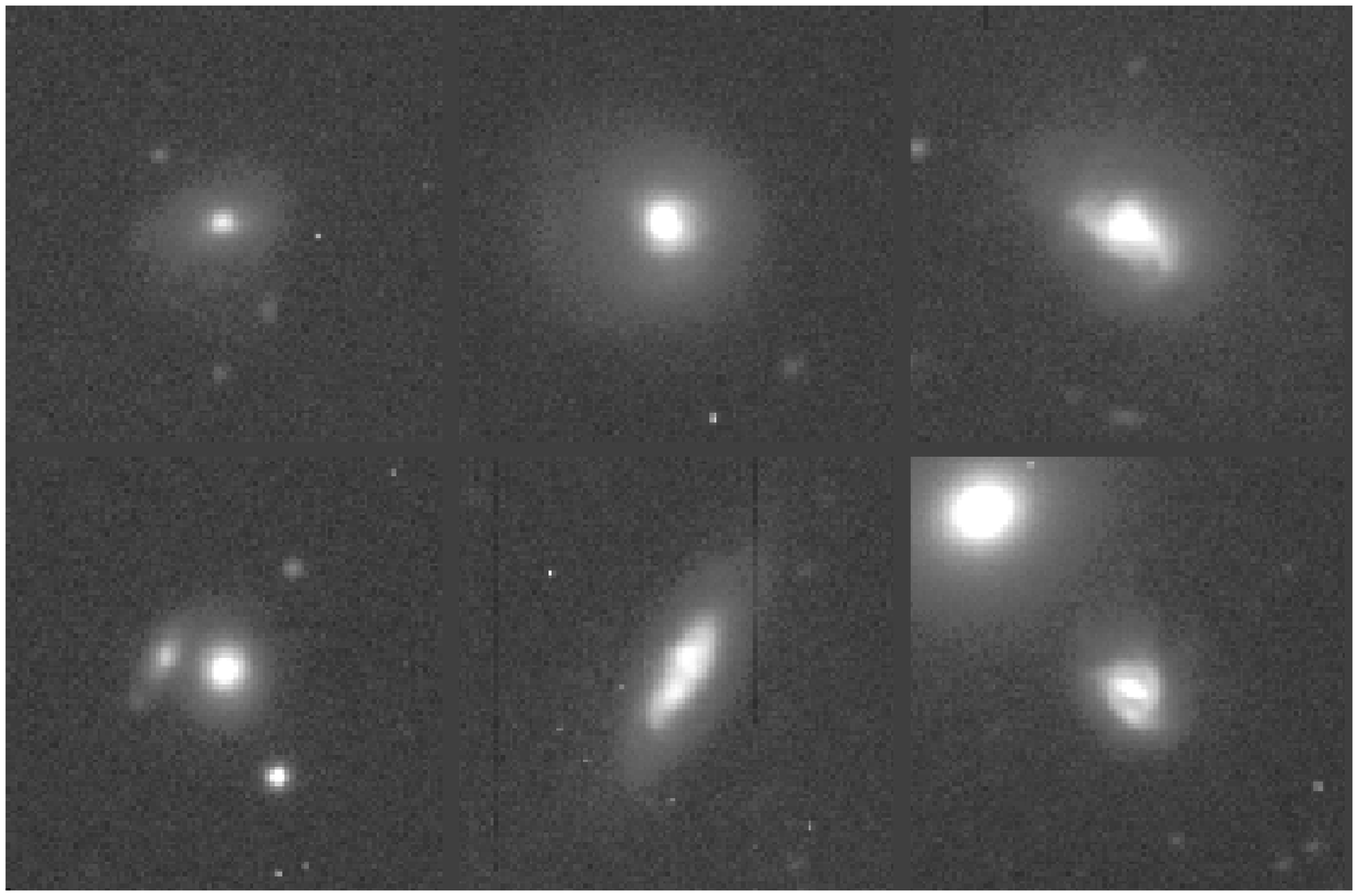}
\caption{V bandpass images of four of the BD86 spirals are compared with
images of the early-type galaxies Coma D-15 and D-45 (lower left and top left
panels, respectively).  The BD86 spirals are D54 and D169 (middle upper and 
lower panels) and D51 and D195 (upper and lower right panels).
In contrast to the BD86 spirals, the star
formation in D15 and D45 is taking place in what appears to be the bright 
central ``bulges'' of these galaxies.}
\label{bdpics}
\end{figure}

Large errors in morphological classification (viz., confusing a spiral galaxy
for an E or S0)
appear unlikely at the distance of the Coma cluster.  In Fig. \ref{bdpics} we compare
V band images of the two current star forming (CSF) early-type galaxies
D15 and D45 with several of the blue disk galaxies in Coma studied by Bothun
\& Dressler (1986; hereafter BD86).  All images are 400 second exposures taken with the 1.2-m
telescope at the F. L. Whipple Observatory.  As can be seen from Fig. \ref{bdpics},
the disks of D15 and D45 are quiescent in comparison with those of the
BD86 spirals.  However, high resolution images with
WFPC2 on {\it HST} reveals that what appears to be the central ``bulges''
of D15 and D45 are actually bright star-forming regions, unresolved on the
ground-based images (Caldwell, Rose, \& Dendy 1999).  Thus, while the 
``bulges'' of D15 and D45 have in fact
been misunderstood from the ground, the {\it HST} images reinforce the idea
that D15 and D45 have had star formation quite unlike a normal spiral, in
that the star formation is highly centrally concentrated.

On the other hand, two of the best-studied clusters, DC2048-52 and 
DC0326-53/0329-52, are at twice and 2.5 times the distance of Coma,
respectively.  One might question whether at this distance the ground-based
morphological distinctions between early-type and late-type galaxies are
reliable.  If, in fact, the emission-line galaxies in the clusters are simply
misclassified spirals (and thus not unusual at all), then the spatial 
distribution of the star formation should be similar to that in ``normal''
spirals, i.e., spread throughout the disk.  To evaluate this ``misclassified
spiral'' hypothesis, we have used the long-slit spectra described in \S 2,
in comparing the spatial distribution of emission lines in the
cluster galaxies to that in the NFGS sample.

To make a quantitative comparison of the emission line distributions in
field and cluster galaxies, we have defined a
measure for the central concentration of emission lines.  For
most of the galaxies, the emission along the slit is smooth enough that we
simply measure the H$\alpha$ flux in the nuclear pixel and in the pixels on
both sides of the nucleus that are closest to the effective radius, where
$r_{eff}$ is taken from Jansen \etal (2000a).  We define the quantity C/R
to be the ratio of
the H$\alpha$ flux in the central pixel to the average H$\alpha$ flux in the
two pixels on either side of the nucleus at $r_{eff}$.  However, for some
of the NFGS galaxies the emission is so concentrated into discrete knots that
the emission right at $r_{eff}$ can be very low, or undetected.  In those 
cases, we use the alternative ratio C/B, which is the ratio of the nuclear
H$\alpha$ flux to the average flux in the two brightest extranuclear knots.

The cluster galaxies in DC2048-52 and DC0326-53/0329-52 have a typical absolute
blue magnitude of -19 (where we have adopted a Hubble Constant of \hub = 70
\kms Mpc$^{-1}$).  The NFGS, by design, covers a very large range in absolute 
magnitude.  To avoid the behavior exhibited by the extreme
bright and faint ends of the  galaxy luminosity function, we restrict our
analysis of the NFGS to the magnitude range -17.4$>$M$_B$$>$-22.0.  In
addition, since the emission lines in our cluster galaxies have characteristic
HII region line intensity ratios, thus indicating star formation rather
than an AGN, we avoid those galaxies in the NFGS whose nuclear spectra are
dominated by an AGN (whether Seyfert or LINER).  
In the one case among the cluster galaxies in which an AGN spectrum dominates 
in the nucleus (D15 in Coma), we measure only the narrow-line
component of the H$\alpha$ emission.  Altogether, we measure C/R and C/B
indices for 66 galaxies in the NFGS, excluding the 33 that are dominated by an
AGN spectrum.  The remainder of the sample falls outside the absolute blue
magnitude limits, or has unmeasureably low H$\alpha$ emission.  We also
measure the same indices in our sample of 5 galaxies in DC2048-52,
3 galaxies in DC0326-53/DC0329-52, and two galaxies in Coma.  Results for the
66 NFGS galaxies and for the 10 cluster galaxies are given in Tables 
\ref{field_table} and \ref{cluster_table}
respectively, where the morphological type, absolute blue magnitude and 
effective radius is listed
for each galaxy, along with the C/R or C/B index.  The default is the C/R
index, i.e., the C/B index is only listed when the emission is highly
resolved into knots.

\begin{figure}
\plotone{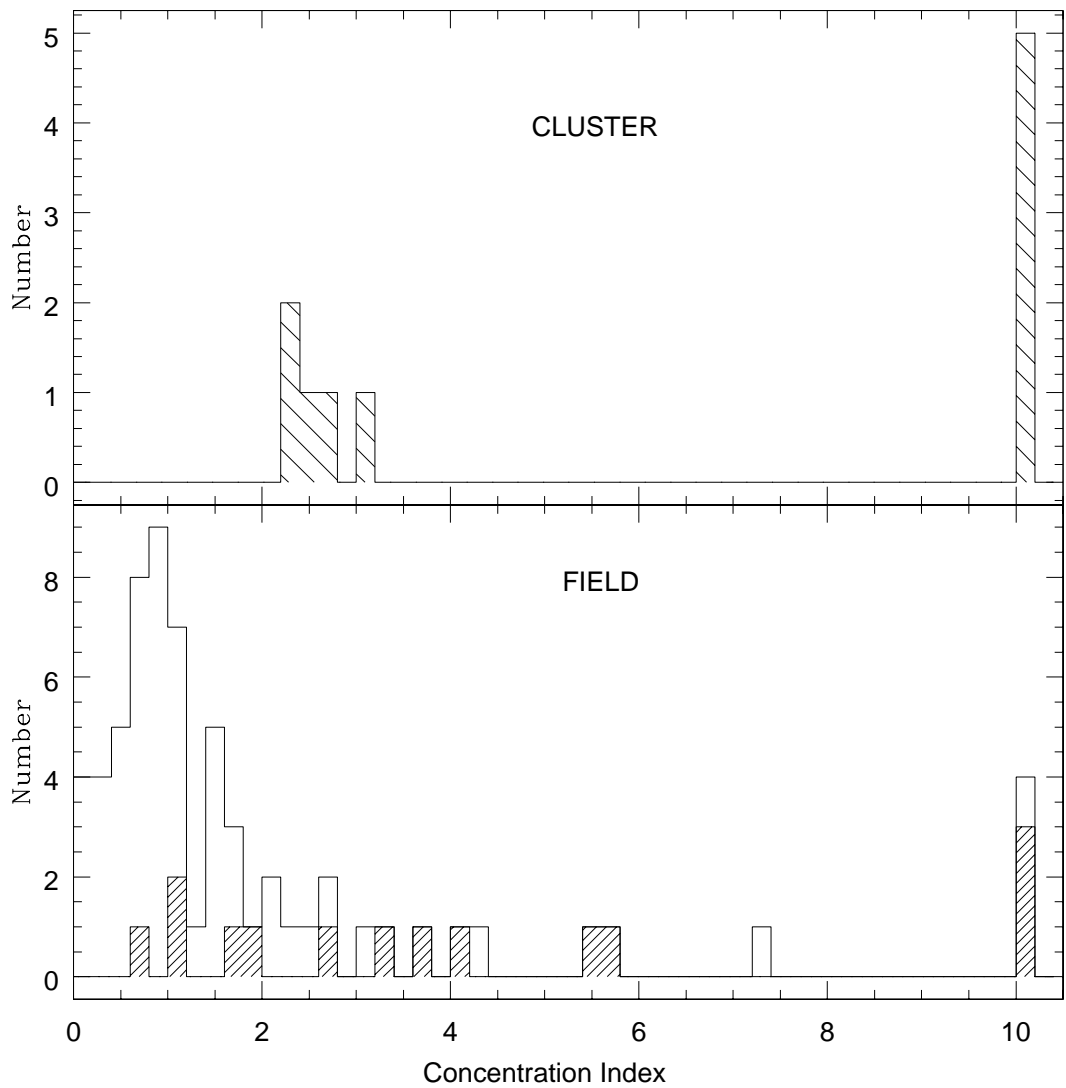}
\caption{The histogram of the C/R ratios of the cluster emission-line
early-type galaxies (top) is compared with that of the NFGS galaxies (bottom).
The shaded part of the NFGS histogram represents those galaxies which have been
classified as Markarian, interacting, starburst, etc.
}
\label{fig_coverr}
\end{figure}

To compare the central concentration of the emission in the cluster versus
the NFGS galaxies, we have plotted histograms of the C/R indices for both the
cluster sample and for the subsample of the NFGS described above.
We again emphasize that the NFGS galaxies satisfying these criteria
consist almost entirely of spirals, since early-type galaxies which have
emission are almost exclusively of the AGN variety.  Since the point of our
comparison is to assess whether the star-forming cluster galaxies might be
misclassified spirals, it is in fact appropriate that the restricted NFGS be 
largely comprised of spirals.
The two histograms are compared in Fig. \ref{fig_coverr},
where the upper histogram is for the cluster galaxies.  For the NFGS galaxies
in the lower histogram we have shaded those galaxies which have been 
classified (based on information given in NED) as Markarian, starburst,
interacting, double, or blue compact dwarf.
Note that for both histograms, we have binned all values of C/R $>$ 10
into a single bin at 10.  The two distributions
are clearly different, in that the majority of the NFGS galaxies have C/R$<$
2.0, while all of the cluster galaxies have C/R $>$ 2.0.  On the other hand,
the distribution of shaded galaxies in the NFGS histogram is quite similar to 
that of the cluster galaxies.  To quantify these assertions we have applied the
Kolmagorov-Smirnoff two-sample test.  In particular, the likelihood that the 
cluster sample and the NFGS are drawn from the same parent population is
only $9.4 \times 10^{-5}$.  On the other hand, the likelihood that the shaded
NFGS and the cluster sample are drawn from the same parent sample is 0.45, while
the likelihood that the shaded NFGS and the remainder of the NFGS are drawn from
the same population is only $5.1 \times 10^{-4}$.
In short, the starforming early-type cluster galaxies 
have more centrally concentrated emission than do typical field galaxies.

\begin{figure}
\plotone{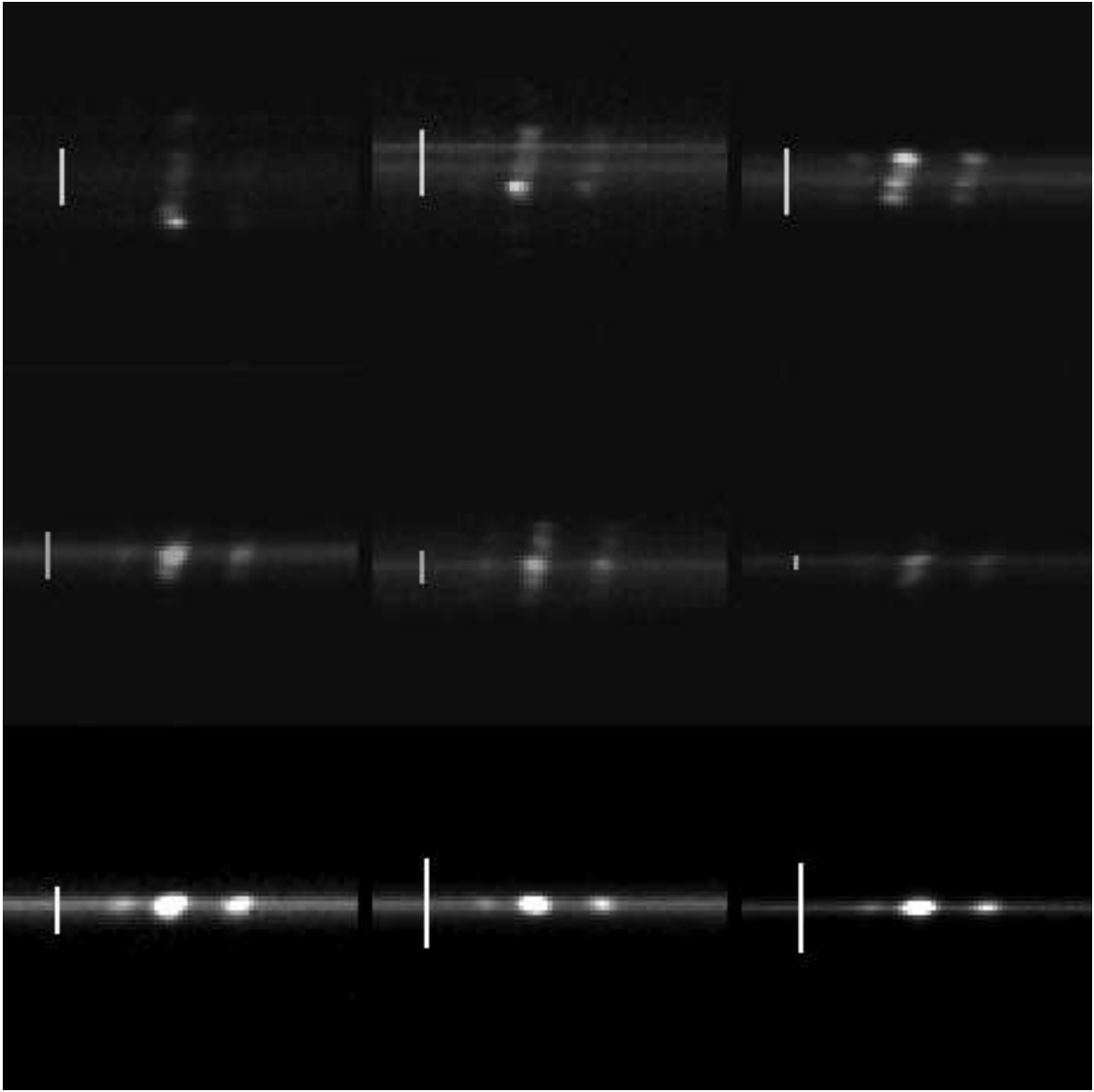}
\caption{Mosaic of long-slit spectra of NFGS galaxies.  Top panels contain
representative galaxies with low central 
concentration emission lines, i.e., with C/R$<$2. Middle panels contain
galaxies with intermediate central concentration emission lines, i.e., with 
2$\le$R$\le$4. The bottom panels show galaxies with high central
concentration emission, i.e., with R$>$4. The fiducial line for each
spectrum is 4 kpc long.}
\label{fmosall}
\end{figure}

\begin{figure}
\plotone{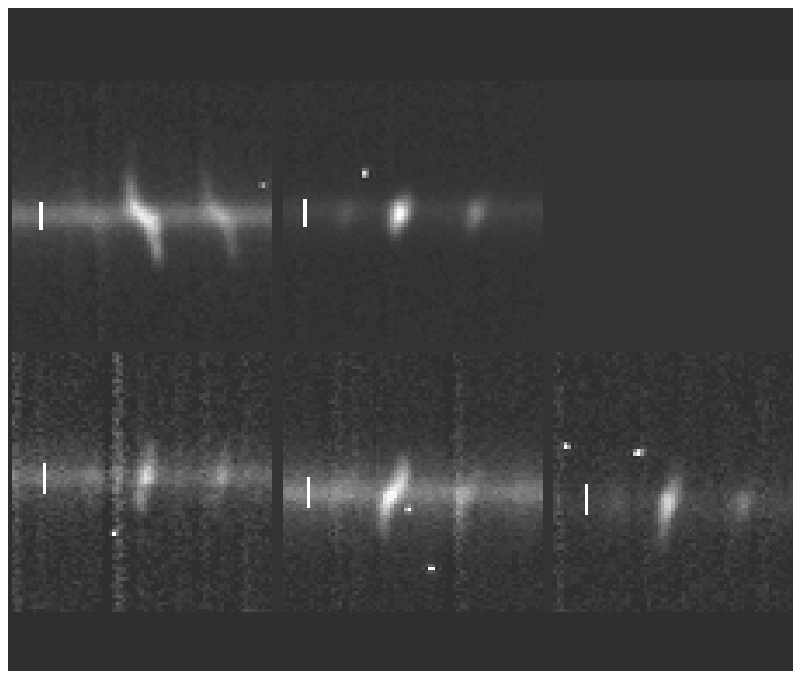}
\caption{Mosaic of long-slit spectra of cluster galaxies with intermediate
concentration emission lines. The fiducial line for each spectrum is 4 kpc
long.}
\label{cmosmid}
\end{figure}

\begin{figure}
\plotone{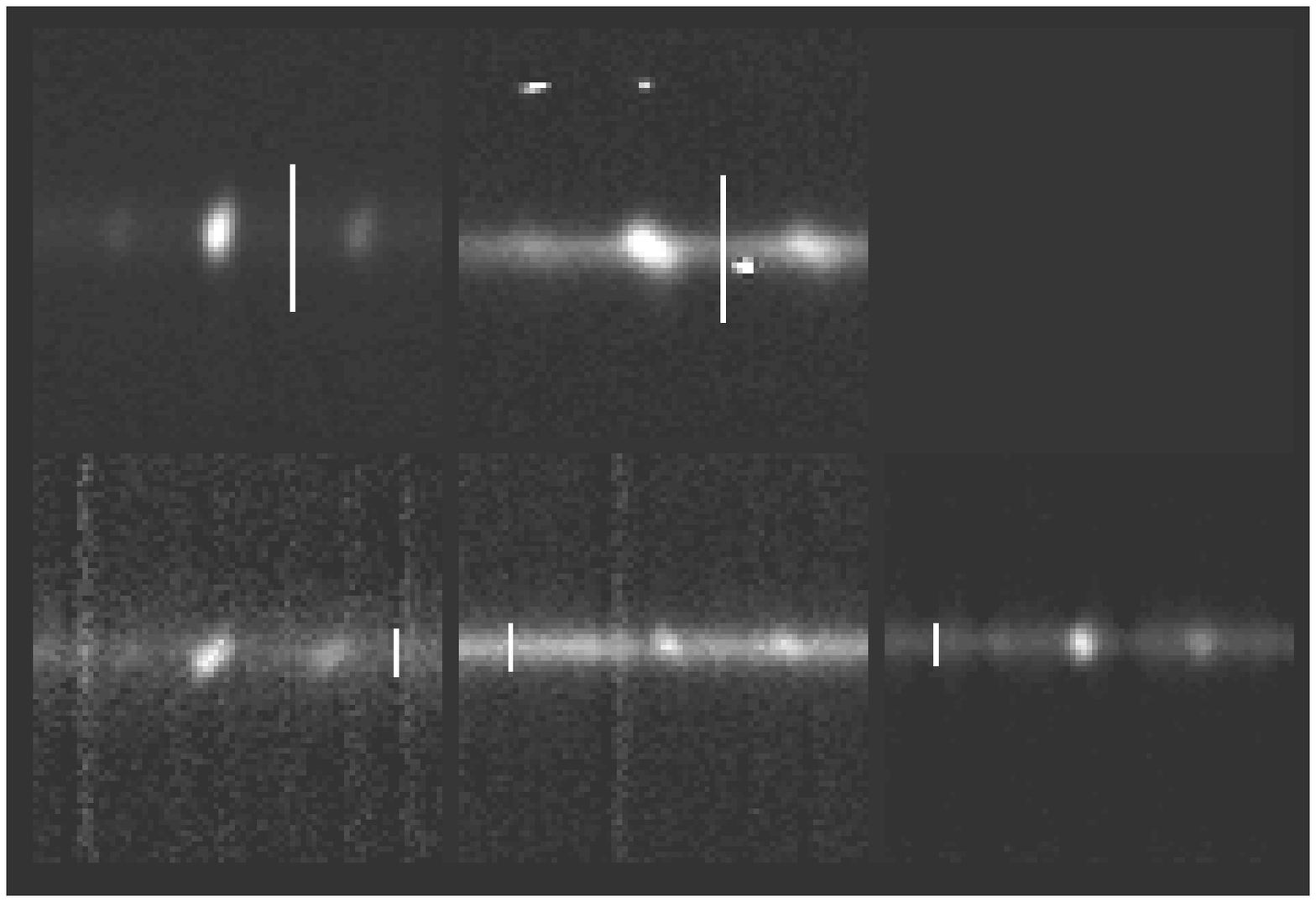}
\caption{Mosaic of long-slit spectra of cluster galaxies with high
concentration emission lines. The fiducial line for each spectrum is 4 kpc
long.}
\label{cmoshigh}
\end{figure}

An additional impression of the differences in emission profiles between the
cluster sample and NFGS can be gained from examination of Figs. 3 - 5, where
the long-slit spectra of galaxies with various ranges in C/R ratios are
shown.  Specifically, the NFGS has been divided into intervals of low
concentration (C/R $<$ 2.0), intermediate concentration (2.0$\le$ C/R$\le$ 4.0),
and high concentration (C/R$>$4.0).  Representative spectra of NFGS galaxies
from these
three classes are reproduced in Fig. \ref{fmosall}.
The corresponding intermediate
and high concentration long-slit spectra of the cluster galaxies are presented
in Figs. \ref{cmosmid} and \ref{cmoshigh}; there are no low concentration 
cluster galaxies in our sample of 10 objects.

In conclusion, early-type galaxies in nearby clusters with current star 
formation, as evidenced by emission lines with [NII]/H$\alpha$ characteristic 
of HII regions, have 
unusually centrally concentrated emission, when compared to a sample of 
relatively isolated galaxies.  In addition, many of the small ($\sim$20\%)
fraction of field galaxies which do have emission as centrally concentrated as 
the early-type galaxies in clusters are known as starburst and/or Markarian
or blue compact dwarf galaxies. This further underscores our principle
conclusion that the cluster early-type galaxies are experiencing unusual
star formation, i.e., are not normal spirals that have been 
misclassified as early-type galaxies.  In addition, the unusual (central)
location of the star formation suggests that it has been concentrated into
a burst.

\section{The Pegasus I Cluster and its Unusual Early-Type Galaxies}

We now move on to discuss imaging of four early-type galaxies, with ongoing
or recent
star formation, in a nearby cluster to further assess the nature of
star formation episodes in early-type galaxies.

As briefly mentioned in \S 1, the Pegasus I cluster is a nearby cluster of
galaxies having a low velocity dispersion, no detected x-ray emission, and
no evidence for systematic depletion of HI in its spiral galaxies 
(Giovanelli \& Haynes 1986).  These facts taken together indicate that
ram pressure stripping cannot play an important role in the evolution of 
galaxies in Pegasus I, so that unusual evolutionary events in the cluster
must be attributed to other causes.  Consequently, Pegasus I represents a
relatively
low-density cluster environment for investigating the triggering mechanism
of star formation in cluster galaxies.  This investigation could also shed light
on the presence of field K+A galaxies.

Based on the presence of enhanced Balmer absorption lines and/or emission,
Vigroux \etal (1989) found that four Pegasus I early-type galaxies show
evidence for ongoing or recently completed star formation 
activity.\footnote{There were originally five such galaxies, but 
it later became evident that one of the five unusual galaxies, namely NGC7583,
has been misidentified.  Another galaxy, NGC7604,
which is an asymmetric spiral, is incorrectly called NGC7583 in the
Catalog of Galaxies and Clusters of Galaxies (Zwicky \etal 1961). The
incorrect coordinates are also published in Chincarini \& Rood (1976),
and were used by Vigroux \etal (1989). The real NGC7583 is 
a background galaxy, at a redshift of 12610 \kms, as clarified in NED.  Thus
the strong emission lines and enhanced Balmer absorption reported by
Vigroux \etal (1989) for NGC7583 are not necessarily unusual, since they 
actually refer
to the spiral galaxy NGC7604 in Pegasus I.}  These four early-type galaxies
have unusual star formation
histories for early-type galaxies, and hence are good candidates for
further study in regard to the mechanism that triggers star formation in 
cluster galaxies.  

To clarify the nature of the recent star formation, we
obtained B and R passband CCD images of the four galaxies.  The 
observations are described in \S 2.  In what follows we address two
primary questions from the Pegasus I imaging data.  First,
what can we deduce about the spatial distribution of the latest star formation
in the four galaxies?  Does it appear to have been widespread, or perhaps
centrally concentrated?  Second, is there evidence for morphological
disturbances in the galaxies which could be associated with the recent star
formation?  The first question is most directly addressed from examination of
the radial color (and luminosity) distributions, while the second question is 
best evaluated from the appearance of the broadband images.  
Here we also address a third
question through the Arecibo 21 cm observations, namely, whether gas is still
resident in the galaxies after the last star formation was completed.

\subsection{Radial Luminosity and Color Profiles}

\begin{figure}
\plotone{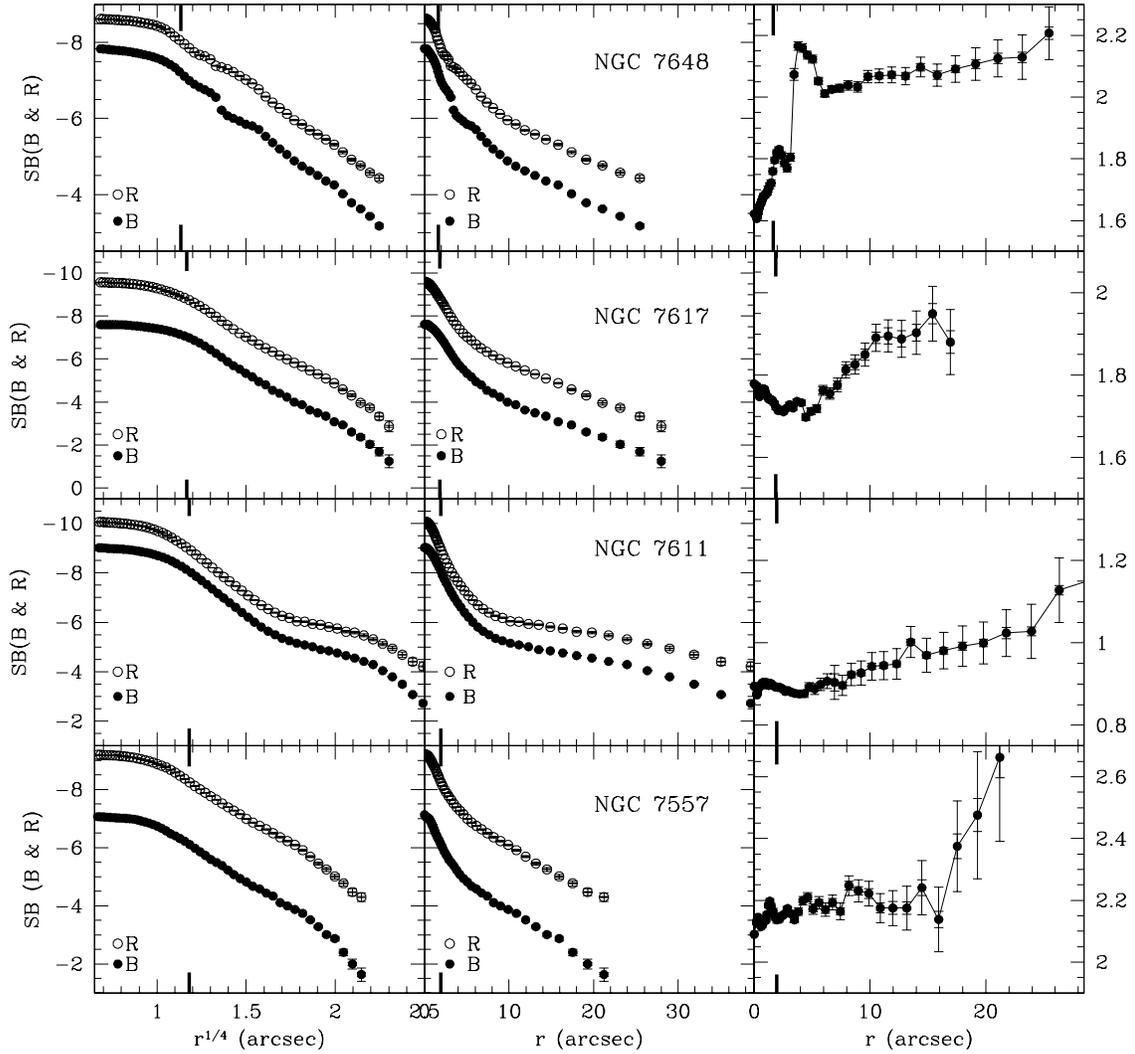}
\caption{Azimuthally averaged radial luminosity and color profiles are plotted
for the four Pegasus I early-type galaxies.  The radial B and R profiles are
plotted versus both r$^{1/4}$ and r in the left and center panels, respectively,
while the $\Delta$($B\!-\!R$) color profiles are shown in the right panels.  The thick solid lines
marked on the horizontal axis indicate the seeing size (FWHM).  The two sets of
error bars indicate the statistical errors and the errors calculated from
estimated uncertainties in sky subtraction (see text); the latter errors
dominate at larger radii.}
\label{fig_profs}
\end{figure}

We begin with an analysis of the radial surface brightness and color profiles for
the four galaxies, which are plotted in Fig. \ref{fig_profs}.
The azimuthally-averaged radial surface brightness profiles are plotted both as a
function of radius, r, and r$^{1/4}$, to aid in assessing the contributions of
disk and bulge.  In the case of NGC7648 a pure r$^{1/4}$ law appears to fit the
observed luminosity profile satisfactorily at all radii, except for a mild
inflection at $\sim$1.5 in r$^{1/4}$, ($\sim$5\arcsc).  However, for the
other three galaxies, departures from the r$^{1/4}$ law are clearly evident in
the outer parts of the profiles, thus indicating that these galaxies have disks.

We now turn to the radial color profiles. Two of the galaxies, NGC7617 and 
NGC7648, show unusual color gradients
in the sense that they become redder with increasing distance from the nucleus.
Such behavior is the reverse of the general blueward color trend with increasing
radius exhibited by most E/S0 galaxies (a typical early-type galaxy of their
luminosity has a color gradient of 
$\Delta$($B\!-\!R$)/$\Delta$(log r)$\sim$-0.1, 
e.g., Vader \etal 1988; Peletier
\etal 1990; Balcells \& Peletier 1994). Specifically, the radial color profile 
of NGC7617 shows a steady
reddening from $\sim$2\arcs (i.e., the seeing limit) to $\sim$15\arcsc, by
slightly more than 0.2 mag in $B\!-\!R$, while in NGC7648 the nucleus is 
0.5 mags bluer than at a radius of 10\arcsc.  
In NGC7617 the enhanced Balmer absorption lines are
accompanied by [OII]$\lambda$3727 emission, indicating that star formation is
still ongoing.  In the case of NGC7648, the emission is considerably stronger,
indicating that this galaxy is in the midst of a star formation 
episode.  In contrast, NGC7557 and
NGC7611 have flat radial color distributions within the uncertainties; these 
two galaxies have true ``K+A'' spectra
in the sense that enhanced Balmer absorption lines are seen, but no detectable
emission. 
To summarize, a very blue central region is seen in the case where a
starburst is in progress (NGC7648), a less dramatically blue nucleus is 
evident in the case where ongoing star formation is less pronounced (NGC7617),
and no color gradient is present in the two cases where there has been recent,
but no longer ongoing, star formation (NGC7557 and NGC7611).

\subsection{Morphologies}

\begin{figure}
\plotone{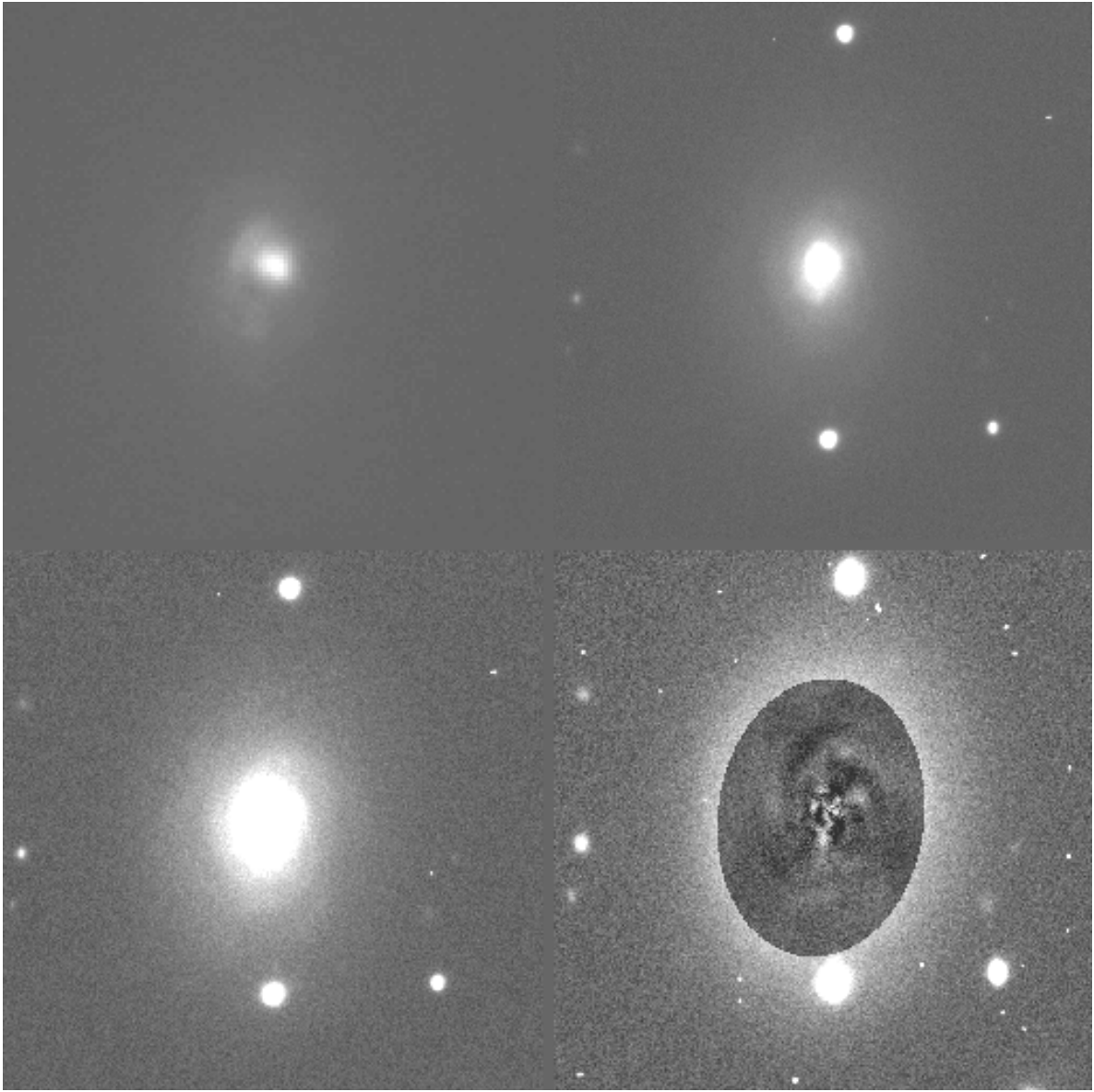}
\caption{Mosaic of images of NGC7648.  Top left (a) is a B band image showing
only the structure in the central region.  It is displayed at twice the
scale of (b) - (d).  Top right (b) is an R band image, at intermediate
contrast.  The separation of the two bright stars is $\sim$80\arcs.  Bottom
left (c) is the same R band image, at higher contrast to emphasize fainter
features.  Bottom right (d) is the model-subtracted version of the R band
image, which emphasizes fine structure at all brightness levels. In all four
panels, North is to the left, and East is to the bottom. }
\label{n7648_mos}
\end{figure}

The morphologies of the four galaxies provide an additional perspective.  
While the two true ``K+A'' galaxies NGC7557 and NGC7611 show little 
morphological peculiarity, NGC7617 has a long curving dust lane, and
NGC7648 shows a variety of interesting features.  In Fig. \ref{n7648_mos}(a), which is a
B bandpass image of NGC7648, an asymmetric nucleus is visible, along with
a luminous arc centered 2.5\arcs to the NW of the galaxy center.
A second arc curves from the eastern end of the first arc to a knot 
6\arcs E of the nucleus. The second arc is also visible in 
Fig. \ref{n7648_mos}(b), which is taken from an R band image, at a factor of
two smaller scale.  At lower surface brightness levels,
i.e., outside 10\arcs in radius,
faint ripples can be discerned.  The faintest ripples are
better seen in Fig. \ref{n7648_mos}(c), which is the same R band image
displayed at a different contrast level.  The fine structure features 
are perhaps best displayed in the difference image between the sky-subtracted
frame and the model frame created from the ellipse fitting (cf., \S 2), which 
is shown in Fig. \ref{n7648_mos}(d). The asymmetric morphological
features discussed above are characteristic of disturbances
found in the late stages of a major (i.e., roughly equal-mass) merger, such as
NGC3921 (Schweizer 1996 and private communication).

\begin{figure}
\plotone{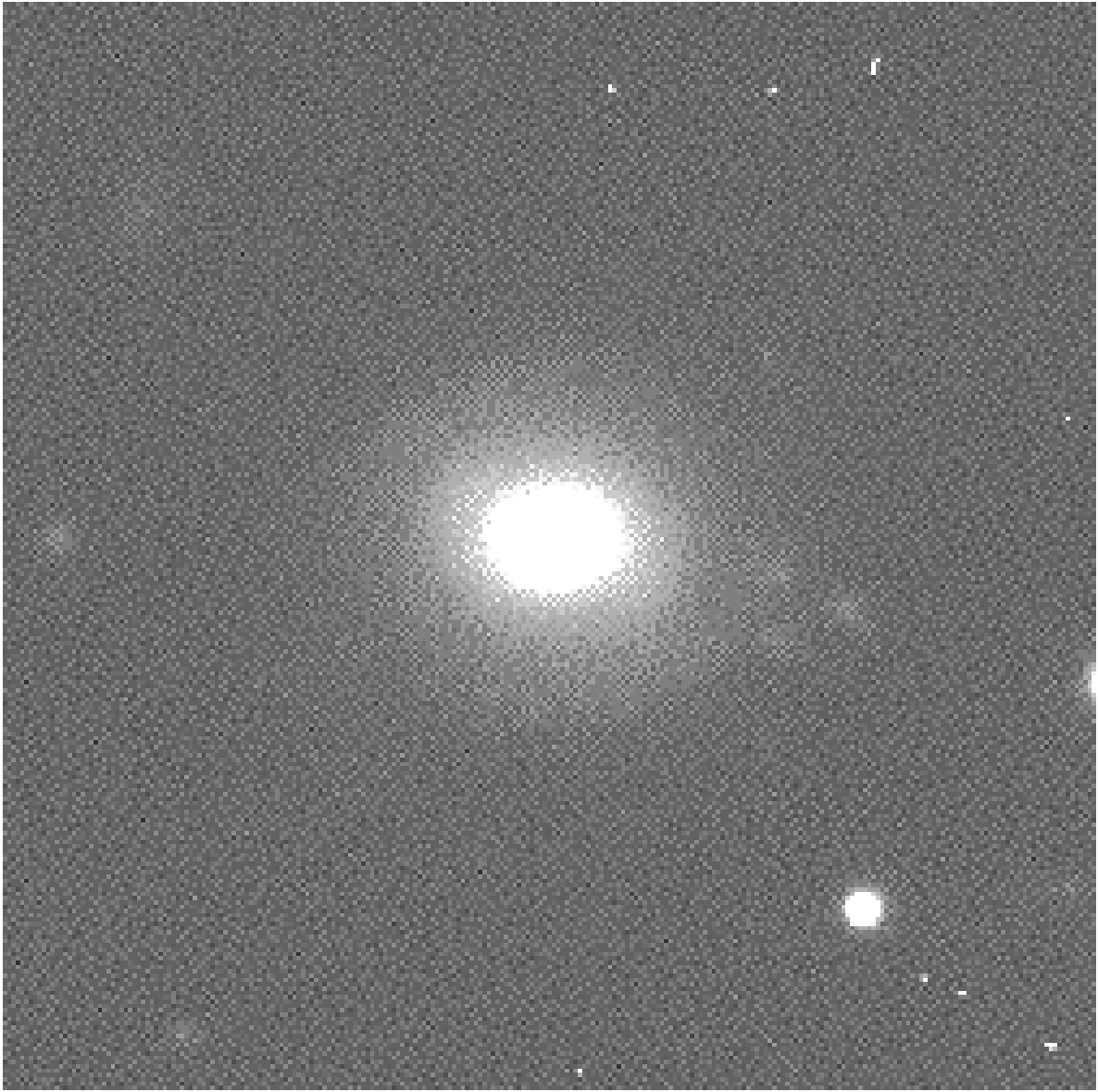}
\caption{R band image of NGC7557. North is to the left, and East is to the
bottom.}
\label{n7557_fin}
\end{figure}

\begin{figure}
\plotone{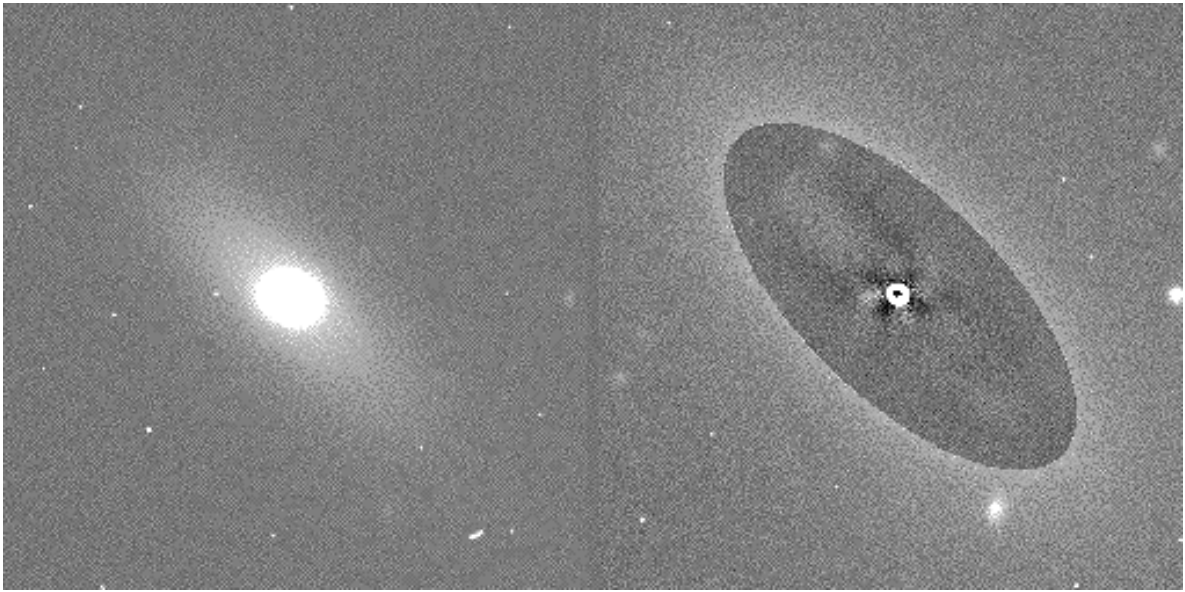}
\caption{R band images of NGC7611. On the left (a) is a sky subtracted
image.  On the right (b) is the ``fine structure'' image for a deeper
exposure, for which the few central pixels are saturated, thereby creating
a central artifact.  The bisymmetric features are seen just outside the
nucleus.  North is to the left, and East is to the bottom.}
\label{n7611_mos}
\end{figure}

\begin{figure}
\plotone{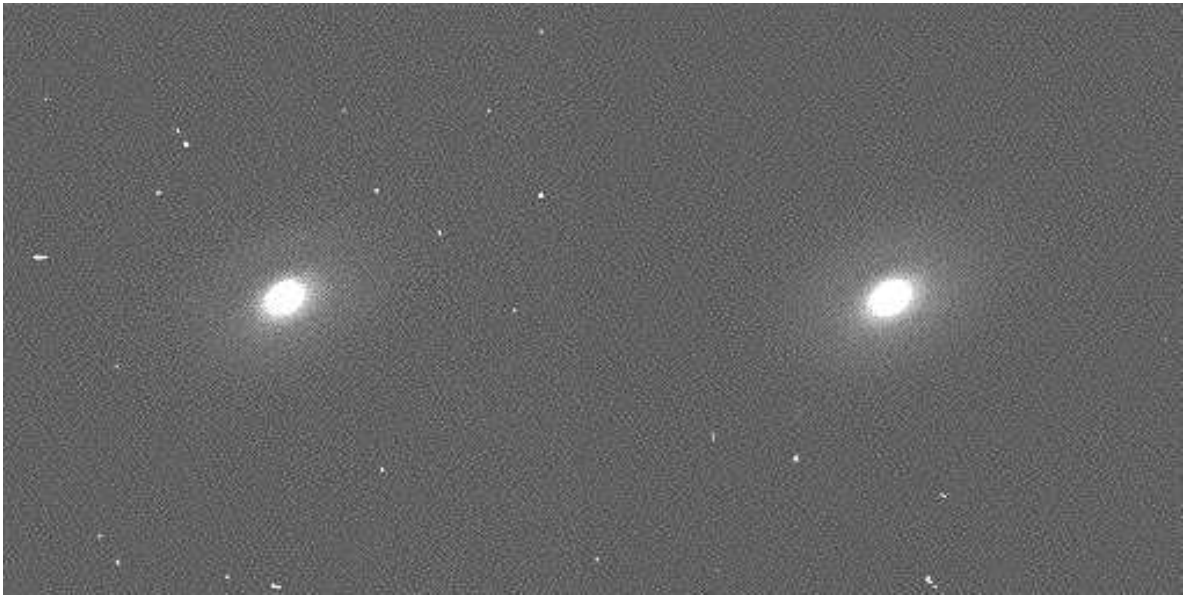}
\caption{Left (a) and right (b) are the B and R band images of NGC7617.  The
curving dust lane can be seen to the north and west of the galaxy.
North is to the left, and East is to the bottom.}
\label{n7617_mos}
\end{figure}

In contrast, both NGC7557 and NGC7611 show only minor morphological
peculiarities.  There is faint outer spiral structure in
NGC7557, as seen in Fig. \ref{n7557_fin}. NGC7611 has a bright
bulge and a faint disk, which are seen in \ref {n7611_mos}(a).  In the ``fine
structure'' image, obtained by subtracting the model fit from the original
R band image, it can be seen (in \ref {n7611_mos}(b)) that
elliptical isophotes provide a poor fit to the light distribution inside the
central $\sim$3\arcsc, where a number of darker and brighter regions can
be seen.  The centermost bright and dark features are artifacts caused by the
few saturated pixels in this image.  However, the saturated pixels do not
affect the model fitting, and the isophotal deviations 
are apparent as well on the model-subtracted unsaturated image 
in \ref {n7611_mos}(a), only are not as visible as in the better exposed image.
The bilateral symmetry of the isophotal deviations suggests a
boxy structure in the nucleus, which is not uncommon in early-type
galaxies (e.g., Bender, D\"{o}bereiner, \& M\"{o}llenhoff 1988).  In NGC7617 
the above-mentioned dust
lane, which extends for about 45\dug from NW to N, can be seen in both B and R
images, which are displayed in Fig. \ref{n7617_mos}(a) and (b).

\subsection{21 cm Results}

\begin{figure}
\plotone{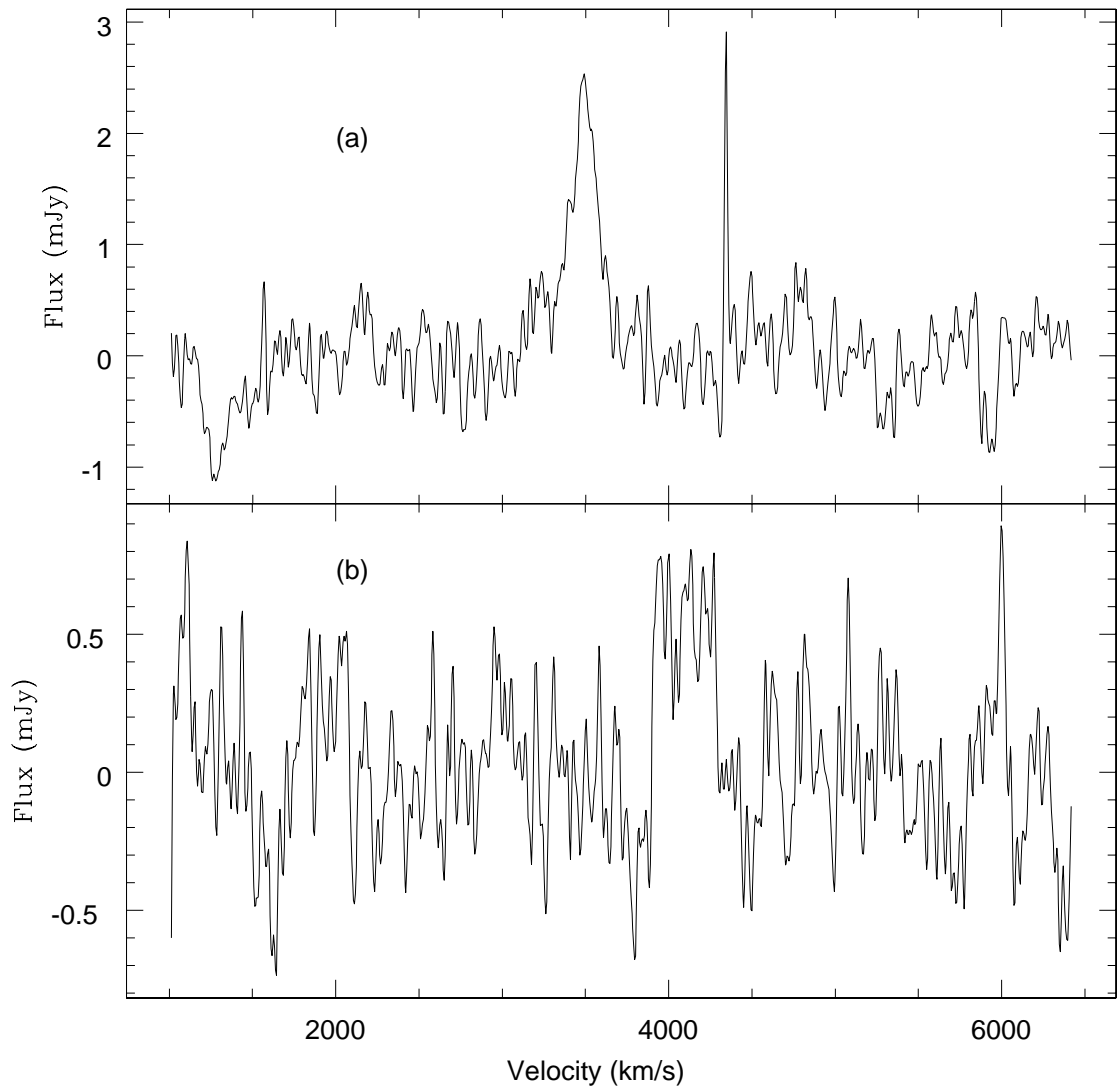}
\caption{21 cm spectral observations of (a) NGC7648 and (b) NGC7617 obtained 
with the Arecibo radiotelescope.}
\label{fig_HI}
\end{figure}

The HI content of the Pegasus galaxies should also give us clues as to the origin of
their most recent (or ongoing) star formation.  NGC7648 and NGC7617 are 
both detected in HI at their previously known optical velocities, at the 10$\sigma$ and 6$\sigma$ levels respectively,
while NGC7611 and NGC7557 are not.  The continuum-removed
spectra for NGC7648 and NGC7617 are plotted in Fig. \ref{fig_HI}.  
Assuming a distance of 60 Mpc to Pegasus I,
we find a 3$\sigma$ upper limit of 1 x $10^8 M_{\sun}$ and 2 x $10^8 M_{\sun}$ 
for NGC7611 and NGC7557 respectively (cf.
\S 2 for a discussion of how the upper limits were determined).
On the other hand, the total HI masses in NGC7648 and NGC7617 are 
$\sim$4 x $10^8 M_{\sun}$ and 2 x $10^8 M_{\sun}$ respectively.  Data on the HI
detections and upper limits for the Pegasus I galaxies
are summarized in Table \ref{HI_table}.  Upon inspection of
the 21 cm profiles, it is evident that NGC7617 exhibits a flat profile,
characteristic of a rotating disk, while NGC7648 shows a predominantly peaked
profile, which is expected if much of the remaining HI has been funneled into
the central region.  Due to the relatively high levels of time-variable RFI that occurred
during the NGC7648 observations, which were taken near sunrise, it is difficult
to assess the reality of the velocity features on either side of the main
peak in the profile of NGC7648. If real, they indicate the presence of
higher velocity material that may be distributed at larger radius in a
rotating disk.

\subsection{NGC7648}

From the above discussion it is evident that NGC7648 offers a 
promising opportunity to explore the properties of an early-type galaxy in a
nearby cluster that is undergoing a central starburst.  Here we compare the
observed properties of NGC7648 to those of several well-studied galaxies
considered to be prototypical cases of late-stage mergers.  Specifically,
we consider NGC3921 and NGC7252, both of which are considered to be late-stage
mergers of nearly equal mass disk galaxies (Schweizer 1996; 1982), and NGC2782
and NGC4424,
which are suspected to
be late-stage mergers involving an intermediate (4:1) mass ratio (Smith 1994;
Kenney \etal 1996).  

The principal characteristics of the above galaxies are summarized in 
Table \ref{merger_table}, where all observations taken from the literature
have been placed on a common Hubble constant of \hub = 70 \kms Mpc$^{-1}$.
A common theme for these
late stage mergers is the presence of substantial atomic hydrogen in the
{\it outer regions} of the galaxies, i.e., typically associated with the stellar
tidal tails (e.g., Hibbard \etal 1994; Jogee, Kenney, \& Smith 1999).  
In contrast, the molecular
gas tends to be confined to the main bodies of the galaxies 
(Hibbard \etal 1994; Jogee, Kenney, \& Smith 1999), and the ionized gas, as 
represented by H$\alpha$ emission, tends to be very centrally concentrated.
An r$^{1/4}$ profile and prominent tidal tails are evident in roughly equal mass
mergers, while the appearance of an exponential profile and lack of tidal tails
are characteristic of intermediate mass ratio mergers.  In general, the
mergers show concentrated ongoing star formation, as evidenced by strong
Balmer absorption lines (e.g., Fritze-von Alvensleben \& Gerhard 1994 for
NGC7252; Schweizer 1996 for NGC3921), and large far-infrared (FIR) luminosities.

NGC7648 does share some of the key characteristics of late-stage mergers,
beyond the morphological knots and ripples discussed in \S 4.2.  In particular,
the FIR luminosity and colors, radio power, and optical emission and absorption
spectrum are all indicative of a burst of star formation.  However, NGC7648 
shows puzzling departures from the prototypical merger cases summarized in
Table \ref{merger_table}.  While no tidal tails are evident in NGC7648, which 
is a characteristic of the two intermediate mass ratio merger candidates, 
current data appears to favor an r$^{1/4}$ law profile (although a deeper image 
is clearly needed).  A pure r$^{1/4}$ law profile is expected
from an equal mass merger, in which violent relaxation has played a strong
role.  Furthermore, while NGC7648 does show an elevated star formation
level, the HI content of the galaxy is very low.  In addition, although there
is no direct information regarding the spatial distribution of the atomic
hydrogen, the peaked HI profile from the Arecibo observations suggests a
centrally concentrated distribution.  Unfortunately, there is no
data concerning the molecular gas content of NGC7648.  Overall, no clearcut
picture emerges as to how to place NGC7648 within the context of late-stage
mergers.  We examine the implications of this comparison in \S 5.3.

\subsection{Summary}

Given the variety of data presented in this section, we briefly summarize the
primary results before going on in \S 5 to examine the implications 
of both the long-slit spectroscopy of \S 3 and the Pegasus I results.
The main conclusion to be drawn from the Pegasus I data is that in two of the
four galaxies with ongoing or recent star formation, the central $\sim$1 kpc in radius
is substantially bluer than the surrounding disk.  Since color gradients in
early-type galaxies tend to run in the opposite direction, we consider the
bluer centers to provide conclusive evidence that the latest
star formation has been centrally concentrated in those two galaxies.
This is especially true for the one case where star formation is still very active, 
NGC7648. {\it Here the evidence is clear that a nuclear starburst is actually in
progress, as opposed to the truncation of disk star formation.} Thus our results
on NGC7617 and NGC7648 in Pegasus I are in accord
with those for the cluster emission-line galaxies in \S 3, 
namely, some galaxies in nearby clusters are currently
undergoing star formation that is completely unlike disk star
formation in a normal spiral galaxy. 
The overall resemblance of the morphological irregularities
found in NGC7648 to that of late stage ``field'' merger systems, as well as the
centralized star formation, 
suggests that NGC7648 is the result of some kind of merger, though other
possibilities remain, as is mentioned in \S 5.3.

The situation regarding the two K+A galaxies, NGC7557 and NGC7611, is unclear because neither of these
galaxies has a blue nuclear region (although we note that a fading centrally
concentrated starburst would rapidly tend not to have a noticeable color gradient).
Long slit spectra showing the Balmer
absorption lines (a more sensitive diagnostic than broadband color
gradients) are needed to sort out whether these two also had centrally
located star formation.  Whatever the case, ram pressure stripping is not
a candidate for causing the unusual star formation history in the Pegasus I galaxies.

\section{Discussion}

To summarize, two principal results have been obtained.  First, long-slit
spectroscopy of emission-line galaxies in the Coma, DC2048-52, and
DC0326-53/DC0329-52 nearby rich clusters has demonstrated that centralized
(in comparison to a field galaxy sample) episodes of star formation occur in
galaxies inhabiting the cluster environment at the present epoch.  Second,
we have found, on the basis of B and R imaging (viz., the presence of
reverse color gradients in their central regions), that centralized star 
formation
is also occurring in galaxies in the Pegasus I cluster.  The key point regarding
the Pegasus I cluster is that it represents a mild cluster environment in
which ram pressure stripping is unlikely to be a factor.  We now 
elaborate on these two results, placing them into context with other 
research on the evolution of galaxies in clusters.

\subsection{Starbursts versus Truncated Star Formation}

A number of recent investigations have addressed the issue of whether star formation
in cluster galaxies primarily ends through rapid truncation or through a
major starburst, a question that has been debated for some time (e.g.,
Newberry, Boroson, \& Kirshner 1990).  Specifically, studies of the
color-magnitude relations of galaxies in relatively distant clusters (e.g.,
Kodama \& Bower 2000), as well as the statistics of emission line
equivalent widths (e.g., Balogh \etal 1997), indicate that many, if not
most, of the galaxies are losing their gas through a truncation
of disk star formation, rather than via a major outburst of star formation.
However,
the evidence presented here, along with the evidence of centralized star
formation in cluster galaxies already published in Caldwell \etal (1996) and
Caldwell, Rose, \& Dendy (1999), demonstrates that in at least some cluster
galaxies, centralized episodes of star formation have taken place which are
very different from ``normal'' star formation in a disk.  Thus some mechanism
must be sought that either drives gas to the central regions of galaxies and
provokes star formation there, or that selectively removes gas from the
outside and then produces a terminal episode in the center.  As is discussed
in \S 5.2 below, in the Pegasus I cluster the evidence points quite 
clearly in the direction of a tidally induced star formation episode rather
than gas removal by ram pressure.  Thus the cluster
environment is capable of producing both truncated star formation 
and enhanced centralized outbreaks as well.  It is natural to suspect, 
therefore, that both evolutionary mechanisms are important in clusters.

To place the evidence for centralized star formation found here within the
context of other work, we mention two other results which may be related.
Koopmann \& Kenney (1998) have emphasized the peculiar
aspects of the morphologies of spirals in the Virgo cluster.  In particular,
they discuss a class of ``truncated'' spirals (which they refer to as
morphological type St), that have nuclear emission and central light
concentrations typical of late-type
spirals, but quiescent disks that are usually associated with early-type
spirals. Hence the Virgo St spirals, which have centrally concentrated emission,
may be analogs to the early-type galaxies with emission that we are finding
in other, more distant, clusters.  
In addition, Moss \& Whittle (1993, 2000) have found large numbers of spiral 
galaxies in 8 nearby clusters which have ``circumnuclear'' (i.e., centrally
concentrated) emission.  Moreover, the spirals with circumnuclear emission
are associated with morphologies indicative of tidal interaction.
Thus it appears likely that our cluster emission-line galaxies have
much in common with those of Moss \& Whittle.

\subsection{Tidal Disturbances versus Ram Pressure Stripping}

Our second result, obtained from the Pegasus I cluster, addresses the 
role of gas stripping versus tidal interaction in driving galaxy
evolution in clusters.  Various
lines of evidence (cf., discussion in \S 1) have convincingly demonstrated the
importance of ram pressure stripping in depleting the gas reservoirs in
galaxies in the rich cluster environment.  The primary observational
evidence for ram pressure stripping {\it in action} comes from studies such as
Gavazzi \etal (1995) and Kenney \& Koopmann (1999), where the radio and optical
morphologies show that star formation is preferentially occurring
in an asymmetric extranuclear arc.  The inference is that star formation is 
induced at the interface between the leading edge of the infalling galaxy
and the hot cluster ICM.  The observational signatures have now been modelled
by Quilis, Moore, \& Bower (2000), whose 3-d gas dynamical simulations also
show the characteristic bow shock effect.  
On the other hand,
in the mild environment of the Pegasus I cluster, ram pressure stripping
cannot plausibly be a factor, and yet two early-type galaxies are found
with ongoing centralized star formation.  In addition, the morphology of
the most clearcut star forming case in Pegasus I, NGC7648, is 
different from those of the three galaxies in A1367 which are believed to be in
dynamical interaction with the cluster ICM.  While the morphology of NGC7648
is complex, there is no sign of a preferential edge (bow shock)
to the star formation.  Rather, non-spiral disturbances are concentrated into 
knots and perhaps ripples, which are suggestive of tidal interaction.  Thus 
our second result is that tidal disturbances do indeed play an important role
in triggering star formation episodes in the cluster environment.  

Evidence for the role of tidal disturbances in clusters has been presented
in several previous studies.  Rubin, Waterman, \& Kenney (1999) have
demonstrated that approximately half of the spiral galaxies in the Virgo
cluster have disturbed rotation curves.  While they argue that the
disturbances are likely to be tidal in nature, they do not distinguish
between galaxy-galaxy and galaxy-(sub)cluster interactions.  A specific
example of an early-type Virgo spiral believed to be a late-stage merger
remnant is NGC4424 (Kenney \etal 1996).
Moss and Whittle (2000) find that a high percentage of the circumnuclear
Emission line galaxies in nearby clusters have perturbed morphologies
indicative of destabilization through a tidal disturbance of some kind.
Numerous studies of more distant
clusters also provide strong evidence for the importance of tidal interactions
and mergers.  Ground-based imaging by Thompson (1988), Lavery \& Henry (1988),
and Lavery, Pierce, \& McClure (1992) revealed numerous cases of interactions
in distant clusters, a result that has been dramatically confirmed by {\it HST}
imaging (e.g., Dressler 1994a,b; Couch \etal 1998 and references therein).

\subsection{Triggering Mechanism for Centralized Star Formation}

Finally, a key question, which remains to be addressed, is what kind of
tidal (or perhaps other) disturbance triggers
the centralized star formation episodes which are clearly evident in the cluster
galaxies studied in this paper.  Due to its presence in a ram-pressure-free
environment, NGC7648 offers a particularly good perspective on this question.
However, the comparison of NGC7648 with four late-stage merger candidates in
\S 4.4 leads to inconclusive results.  With its low (and likely centrally
concentrated) HI content, and apparent lack of an exponential disk component,
NGC7648 fails to fit in with either the equal mass or intermediate mass ratio
merger scenarios.  The one late-stage merger candidate with similarly low HI
content, NGC4424, is resident in the Virgo cluster, and believed to be a
victim of ram pressure stripping (Kenney \etal 1996).  It appears that either
the progenitor galaxies (perhaps low in atomic hydrogen) and/or encounter
geometry must be fundamentally different from the late-stage mergers considered
here, or the merger scenario might be incorrect.  In that case, another type
of tidal disturbance, e.g., ``galaxy harassment'' (Moore \etal 1996; 1998) or
destabilization from the rapidly varying (sub)cluster gravitational potential
(Byrd \& Valtonen 1990; Bekki 1999), could supply the
triggering mechanism.  Unfortunately,
the data presented here is not sufficient to resolve this important issue.

In conclusion, we have shown that some early-type galaxies in nearby 
clusters are experiencing, or have recently experienced, unusual star
formation histories when compared with those in more isolated galaxies.
Specifically, star formation in these galaxies is more centrally 
concentrated than in the disks of isolated spiral galaxies.  Thus the
cluster environment is clearly influencing the evolution of these galaxies.
The cause of this centralized star formation is somewhat less clear.  In
principle it could be due to the preferential removal, by ram-pressure
stripping, of the atomic gas in the galaxy's disk, thereby allowing only for
star formation in the central region.  Alternatively, it may result from
external tidal perturbations, which cause gas to be driven into the central
region of the galaxy.  The fact that in some cases there is good evidence that
the star formation rate has been substantially enhanced in a recent starburst
favors the tidal triggering mechanism.

\acknowledgements

We thank Riccardo Giovanelli for much advice and encouragement in obtaining
and analyzing the 21 cm observations.  We are also grateful to
JoAnn Eder, Karen O'Neil,
and the staff of the Arecibo Observatory for their help with the 21 cm
observations and reductions.  In addition, we thank
Francois Schweizer for helpful comments regarding the 
morphology of NGC7648.  
This research was supported by NSF grant AST-9900720 to the University of North
Carolina.  Travel to Arecibo was partially supported by the National Astronomy 
and Ionosphere Center.

\newpage

\pagestyle{empty}
\setcounter{page}{0}
\vspace*{1.5in}
\begin{deluxetable}{lrrrrrc}
\tablenum{1}
\tablecolumns{6}
\tablewidth{0pc}
\tablecaption{C/R Values for Field Galaxy Sample}
\tablehead{
\colhead{Galaxy ID} & \colhead{T} & \colhead{$V_r$} & \colhead{$M_B$\tablenotemark{1}} &
\colhead{$r_{eff}$} & \colhead{C/R} & \colhead{Comment\tablenotemark{2}}}
\startdata

A00289+0556 & 7 & 2207 & -18.00 &  9.52 &  0.95 &   \nl
A00389-0159 & 1 & 5417 & -20.65 &  11.13 &  4.39 &   \nl
A00442+3224 & 3 & 5079 & -20.34 &  13.08 &  1.73 &  a,b \nl
A01344+2838 & 4 & 7934 & -20.63 &  6.90 &  0.79 &  a \nl
A01346+0438 & 4 & 3257 & -18.54 &  5.08 &  1.65 &   \nl
A02056+1444 & 3 & 4516 & -20.07 &  11.02 &  2.42 &  a \nl
A02257-0134 & 8 & 1795 & -17.91 &  19.32 &  0.54 &  a \nl
A08567+5242 & 3 & 9092 & -21.63 &  16.70 &  $<$0.62 &  a \nl
A09579+0439 & 3 & 4008 & -19.05 &  7.95 &  1.55 &   \nl
A10171+3853 & 9 & 1999 & -17.83 &  7.95 &  0.93 &  a \nl
A10321+4649 & 5 & 3372 & -18.81 &  6.66 &  10.7 &   \nl
A10337+1358 & 6 & 2869 & -18.64 &  11.06 &  0.86 &  a \nl
A10504+0454 & 5 & 5631 & -19.93 &  4.54 &  5.41 &  b \nl
A10592+1652 & 4 & 2830 & -18.25 &  10.83 &  0.76 &  a \nl
A11040+5130 & 5 & 2269 & -18.99 &  28.02 &  0.40 &  a \nl
A11072+1302 & 5 & 12623 & -21.43 &  5.72 &  3.30 &  a,b \nl
A11310+3254 & 3 & 2603 & -18.49 &  9.98 &  0.88 &  a \nl
A11332+3536 & -3 & 1596 & -17.44 &  6.40 &  2.66 &  b \nl
A11372+2012 & 5 & 10891 & -21.65 &  7.58 &  0.15 &  a \nl
A11531+0132 & 5 & 1748 & -18.42 &  33.62 &  1.44 &  a \nl
A12001+6439 & -2 & 1586 & -17.45 &  8.59 &  5.70 &  b \nl
A12167+4938 & 5 & 3720 & -19.28 &  13.13 &  1.46 &  a \nl
A12331+7230 & 3 & 7134 & -20.41 &  9.69 &  2.28 &  a \nl
A13065+5420 & 3 & 2581 & -18.28 &  12.16 &  0.35 &  a \nl
A13194+4232 & 6 & 3478 & -18.93 &  10.18 &  2.74 &  a \nl
A13361+3323 & 9 & 2420 & -18.29 &  14.28 &  10.90 & b \nl
A13422+3526 & 4 & 2570 & -18.17 &  10.58 &  1.17 &  a \nl
A14305+1149 & 5 & 2246 & -18.35 &  11.16 &  0.39 &  a \nl
A15314+6744 & 5 & 6675 & -20.50 &  13.72 &  0.95 &  a \nl
A15523+1645 & 5 & 2290 & -17.49 &  4.89 &  1.05 &  a,b \nl
A22306+0750 & 5 & 2212 & -18.17 &  7.65 &  0.77 &   \nl
A22551+1931 & -2 & 5926 & -18.97 &  4.54 &  4.09 &  b \nl
A23176+1541 & 7 & 4604 & -19.64 &  10.79 &  0.17 &  a \nl
A23542+1633 & 10 & 1997 & -17.98 &  18.04 &  0.76 &  a \nl
IC1100 & 6 & 6758 & -20.72 &  10.04 &  0.78 &  a \nl
IC1124 & 2 & 5346 & -19.85 &  7.30 &  1.30 &  a \nl
IC1776 & 5 & 3486 & -19.39 &  19.72 &  0.70 &  a \nl
IC197 & 4 & 6401 & -20.31 &  7.90 &  0.62 &  a,b \nl
IC2591 & 4 & 6731 & -20.48 &  8.97 &  1.51 &  a \nl
IC746 & 3 & 4989 & -19.53 &  7.80 &  0.54 &   \nl
NGC2780 & 2 & 1916 & -17.66 &  12.11 &  7.23 &   \nl
NGC3009 & 5 & 4682 & -19.46 &  11.37 &  2.18 &   \nl
NGC3326 & 3 & 7972 & -20.91 &  7.49 &  19.06 &  b \nl
NGC3633 & 1 & 2398 & -18.22 &  7.49 &  3.10 &   \nl
NGC4034 & 5 & 2542 & -18.41 &  20.09 &  0.06 &  a \nl
NGC4120 & 5 & 2411 & -18.42 &  15.85 &  1.03 &  a \nl
NGC4141 & 5 & 2098 & -18.75 &  11.25 &  0.93 &  a \nl
NGC4159 & 8 & 1945 & -17.96 &  11.56 &  0.94 &  a \nl
NGC4238 & 5 & 2910 & -18.83 &  13.01 &  1.72 &  a \nl
NGC4961 & 4 & 2545 & -19.06 &  12.68 &  0.90 &  a \nl
NGC5117 & 5 & 2436 & -18.41 &  16.38 &  1.10 &  a \nl
NGC5230 & 5 & 6829 & -21.74 &  22.91 &  2.12 &  a \nl
NGC5267 & 3 & 6021 & -20.57 &  11.49 &  0.60 &  a \nl
NGC5425 & 5 & 2189 & -18.25 &  12.49 &  0.57 &  a \nl
NGC5491 & 5 & 5818 & -20.68 &  11.81 &  0.18 &  a \nl
NGC5541 & 5 & 7804 & -21.29 &  9.01 &  0.35 &  a \nl
NGC5762 & 1 & 1817 & -18.11 &  11.44 &  0.93 &  a \nl
NGC5874 & 4 & 3309 & -19.70 &  26.14 &  0.26 &  a \nl
NGC5875A & 5 & 2645 & -18.37 &  11.10 &  1.03 &  a \nl
NGC5993 & 3 & 9747 & -21.73 &  11.08 &  0.55 &  a \nl
NGC6007 & 4 & 10629 & -21.86 &  14.33 &  1.40 &  a \nl
NGC6131 & 5 & 5241 & -20.13 &  15.84 &  1.18 &  a \nl
NGC695 & 5 & 9855 & -21.76 &  6.33 &  1.86 &  a,b \nl
NGC7328 & 2 & 3051 & -19.37 &  13.23 &  10.81 &   \nl
NGC7620 & 6 & 9811 & -21.93 &  8.42 &  1.17 &  b \nl
\tablenotetext{1}{Based on the total B magnitude given in Jansen \etal (2000a),
and \hub = 70 km s$^{-1}$ Mpc$^{-1}$}
\tablenotetext{2}{a: used the C/B value in place of C/R; b: known to be a 
Markarian, starburst, or
interacting galaxy, etc. from information given in NED}
\enddata
\label{field_table}
\end{deluxetable}

\setcounter{page}{0}
\vspace*{1.5in}
\begin{deluxetable}{lrrrr}
\tablenum{2}
\tablecolumns{5}
\tablewidth{0pc}
\tablecaption{C/R Values for Cluster Early-Type Galaxy Sample}
\tablehead{
\colhead{Galaxy ID} & \colhead{T} & \colhead{$M_B$\tablenotemark{a}} &
\colhead{$r_{eff}$} & \colhead{C/R} }
\startdata

DC2048\#104 & S0/a &  ---- &  3.0\tablenotemark{b} &   13.47 \nl
DC2048\#148 & S0 &  -18.93 &  3.4 &   2.70 \nl
DC2048\#172 & S0/a &  -18.76 &  3.2 &    3.01 \nl
DC2048\#192 & E &  -19.53 &  3.0 &  10.90 \nl
DC2048\#187 & S0 &  -18.85 &  3.8 &    2.56 \nl
DC0326\#82a & E &  -19.18 &  3.0 &   26.96 \nl
DC0326\#101a & S0/a &  -19.32 &  2.8 &    2.26 \nl
DC0326\#80b & E &  -18.92 &  2.0 &    2.37 \nl
Coma-D45 & Sa &  -18.28 &  4.0 &  106.10 \nl
Coma-D15 & S0 &  -18.97 &  4.0 &   24.10 \nl

\tablenotetext{a}{Based on the total B magnitude given in Caldwell \etal (1993),
or Caldwell \& Rose (1997), and \hub = 70 km s$^{-1}$ Mpc$^{-1}$}
\tablenotetext{b}{No data available -- assumed a value of 3.0, based on other
galaxies in the cluster}
\enddata
\label{cluster_table}
\end{deluxetable}

\setcounter{page}{0}
\vspace*{1.5in}
\begin{deluxetable}{lrrrrr}
\tablenum{3}
\tablecolumns{6}
\tablewidth{0pc}
\tablecaption{21 cm Data for Pegasus I and Coma Galaxies}
\tablehead{
\colhead{Galaxy ID} & \colhead{Detected Flux} & 
\colhead{3$\sigma$ Noise} & \colhead{M$_{HI}$} &
\colhead{$M_B$} & \colhead{M$_{HI}$/$L_B$} \\
\colhead{} & \colhead{(Jy km/s)} & \colhead{(Jy km/s)} & 
\colhead{($M_{\sun}$)} & \colhead{} & \colhead{($M_{\sun}$/$L_{\sun}$)} }
\startdata

NGC7557 & \nodata & 0.27 & $<$2.0 x 10$^8$ & -18.87 & $<$4.3 x 10$^{-2}$ \nl
NGC7611 & \nodata & 0.15 & $<$1.3 x 10$^8$ & -20.50 & $<$5.3 x 10$^{-3}$ \nl
NGC7617 & 0.21    & 0.10 & 1.7 x 10$^8$ &    -19.24 & 2.3 x 10$^{-2}$ \nl
NGC7648 & 0.46\tablenotemark{a} & 0.14 & 3.9 x 10$^8$ & -20.25 & 2.0 x 10$^{-2}$ \nl
\nodata & 0.56\tablenotemark{a} & 0.14 & 4.8 x 10$^8$ & -20.25 & 2.5 x 10$^{-2}$ \nl
Coma-D15 & 0.088 & 0.080 & 2.1 x 10$^8$ & -18.97 & 3.6 x 10$^{-2}$ \nl
Coma-D16 & \nodata & 0.080 & $<$1.9 x 10$^8$ & -18.86 & $<$3.6 x 10$^{-2}$ \nl
Coma-D45 & \nodata & 0.079 & $<$1.8 x 10$^8$ & -18.28 & $<$5.8 x 10$^{-2}$ \nl
Coma-D100 & 0.043\tablenotemark{b} & 0.076 & 1.0 x 10$^8$ &  -19: & 1.7 x 10$^{-2}$ \nl
\nodata  & \nodata\tablenotemark{b} & 0.076 & $<$1.8 10$^8$ & -19: & $<$3.0 x 10$^{-2}$ \nl
\tablenotetext{a}{The lower and upper values correspond to when the wings on
the 21 cm profile are not and are included, respectively.}
\tablenotetext{b}{The first line for Coma-D100 corresponds to results from an
assumed 1.7$\sigma$ detection; the second line corresponds to the 3$\sigma$ 
upper limit.}
\enddata
\label{HI_table}
\end{deluxetable}

\setcounter{page}{0}
\vspace*{1.5in}
\pagestyle{empty}
\begin{deluxetable}{lcccccccc}
\tablenum{4}
\tablecolumns{9}
\tablewidth{0pc}
\tablecaption{Physical Parameters for Merging Galaxies\tablenotemark{a}}
\tablehead{
\colhead{Galaxy ID} & \colhead{Tidal} & \colhead{Exponential} & 
\colhead{M$_{HI}$/$L_B$} &
\colhead{log} & \colhead{log} &
\colhead{log} & \colhead{log} & \colhead{log P(1.415 GHz)} \\
\colhead{} & \colhead{Tails?} & \colhead{Disk?} &
\colhead{($M_{\sun}$/$L_{\sun}$)} & \colhead{[L(FIR)/$L_{\sun}$]\tablenotemark{b}} &
\colhead{f12/f25} & \colhead{f25/f60} & \colhead{f60/f100} &
\colhead{[W/Hz]} }
\startdata
NGC3921 & y & n & 0.09 & \nodata & \nodata & -0.88 & \nodata & \nodata \nl
NGC7252 & y & n & 0.07 & 10.65 & -0.26 & -0.96 & -0.25 & \nodata \nl
NGC2782 & y & y & 0.13 & 10.84 & -0.30 & -0.79 & -0.23 & 21.74 \nl
NGC4424 & n & y & 0.03 & 9.38 & -0.30 & -0.96 & -0.25 & $<$20.25 \nl
NGC7648 & n & n & 0.02 & 10.47 & -0.41 & -0.95 & -0.20 & 21.62 \nl
\tablenotetext{a}{all values taken from the literature have been rescaled to
\hub = 70 \kms Mpc$^{-1}$}
\tablenotetext{b}{The far-infrared (FIR) luminosity is as defined in Table 5 
of Bicay \etal (1995)}
\enddata
\label{merger_table}
\end{deluxetable}

\end{document}